\documentclass{ws-ijmpe}
\usepackage{amsmath}

\begin{document}

\title{OPEN PROBLEMS IN HYDRODYNAMICAL APPROACH \\
TO RELATIVISTIC HEAVY ION COLLISIONS}
\author{{\footnotesize T. Kodama, T. Koide, G. S. Denicol and Ph. Mota%
\thanks{%
Univ. Federal do Rio de Janeiro, Rio de Janeiro, Brazil}\ \ \ } }
\maketitle

\begin{abstract}
We discuss some open problems in hydrodynamical approach to the relativistic
heavy ion collisions. In particular, we propose a new, very simple
alternative approach to the relativistic dissipative hydrodynamics of Israel
and Stewart.
\end{abstract}

\markboth{T. Kodama et al}{Open Problems in Hydrodynamical Apprach to
Relativistic Heavy Ion Collisions )}

\catchline{}{}{}{}{} 

\address{Insituto de Fisica, Uniiversidade Federal do Rio de Janieoro\\
Rio de Janeiro,
Brazil\footnote{State completely without abbreviations, the
affiliation and mailing address, including country. Typeset in 8~pt
Times italic.}\\
first\_author@university.edu}

\begin{history}
\received{(received date)}
\revised{(revised date)}
\end{history}

\section{Introduction}

The phase transitions in strongly interacting bulk matter predicted by the
quantum chromodynamics (QCD) should manifest their existence in different
physical scenarios such as the evolution of inhomogeneities in the early
universe, structure of compact stars, spectra of particles from nuclear
collisions at ultra-relativistic energies, etc. In particular, the
relativistic heavy ion collisions are the unique possibility of observing
such phase transitions in laboratories, permitting us to extract the bulk
properties of the strongly interacting matter at extremely high temperature
and energy density.

As we see from many talks in this conference, the over-all \textquotedblleft
picture\textquotedblright\ of the new states of the matter at extreme
condition achieved in the laboratories (mainly from SPS and RHIC) is now
being configured after more than two decades when the first project on
relativistic heavy ion physics started\cite{Laszlo}. Yet the nature of the
transition from the hadronic phase to the QGP phase is still to be clarified
quantitatively. Of course, in addition to the experimental data, advances in
theoretical studies such as lattice QCD calculations also enriched
understandings of the properties of the strong interacting matter\cite%
{Karsch,fodor}.

Among many signals of QGP, the one which brought a new insight is the
emergence of collective flow in the final state of exploding particles. In
the non-central collisions, an asymmetric distribution of energy density is
created in the first instant of the collision. In the hydrodynamical image,
the driving force to expand the system is proportional to the pressure
gradient, so that this asymmetry with respect to the direction of the
largest pressure gradient should increase as a function of the transverse
momenta of particles. Quantitatively, such anisotropy can be expressed in
terms of the coefficients of Fourier series of the azimuthal distribution of
particles. We define the elliptic flow for a given transverse momentum
window as%
\begin{equation*}
v^{\left( 2\right) }\left( p_{T}\right) =\frac{\left\langle \int d\phi \ %
\left[ d^{2}N/dp_{_{T}}^{2}d\phi \right] \cos 2\left( \phi -\phi _{\vec{b}%
}\right) \right\rangle }{\left\langle \int d\phi \ \left[
d^{2}N/dp_{_{T}}^{2}d\phi \right] \right\rangle },
\end{equation*}%
where $p_{_{T}}$ is the transverse momentum, and $\phi $ and $\phi _{\vec{b}%
} $ are, respectively, the azimuthal angles of the particle and of the
impact parameter vector with respect to a some space-fixed coordinate
system. In practice, the determination of these coefficients from the
experimental data is not trivial since the reaction plane is not given a
priori. In the hydrodynamical calculations, of course the reaction plane is
given from the beginning.

The finite positive value of this coefficient $v_{2}$ is referred to as the
elliptic flow and, from the point of view of hydrodynamics, it is sensitive
to the initial pressure gradient of the system. The above expected behavior
of $v_{2}$ was clearly observed in RHIC data, and the values of observed
values approach to those calculated from the ideal hydrodynamics. At the
same time, for lower (SPS) energies, the experimental values are lower than
the ideal fluid values (See Figs.1 and 2). This fact was interpreted as the
emergence of the new state of the matter which flows almost as an ideal
fluid, while in the hadronic phase the matter suffers from the collisional
viscosity\cite{Miklos}.

\begin{figure}[tbph]
\begin{center}
\includegraphics[width=8cm]{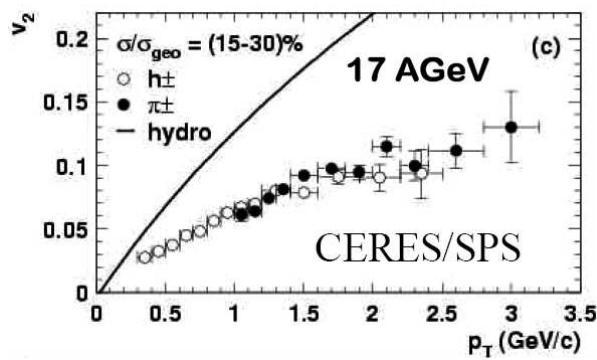}
\end{center}
\caption{Elliptic flow at SPS energies as function of transverse momentum,
compared to the simple ideal fluid hydrodynamical calculation. The observed
values of elliptic flow are below the hydro values. Figure adapted from the
presentation of M. Gyulassy, see this volume}
\end{figure}

\begin{figure}[tbph]
\begin{center}
\includegraphics[width=8cm]{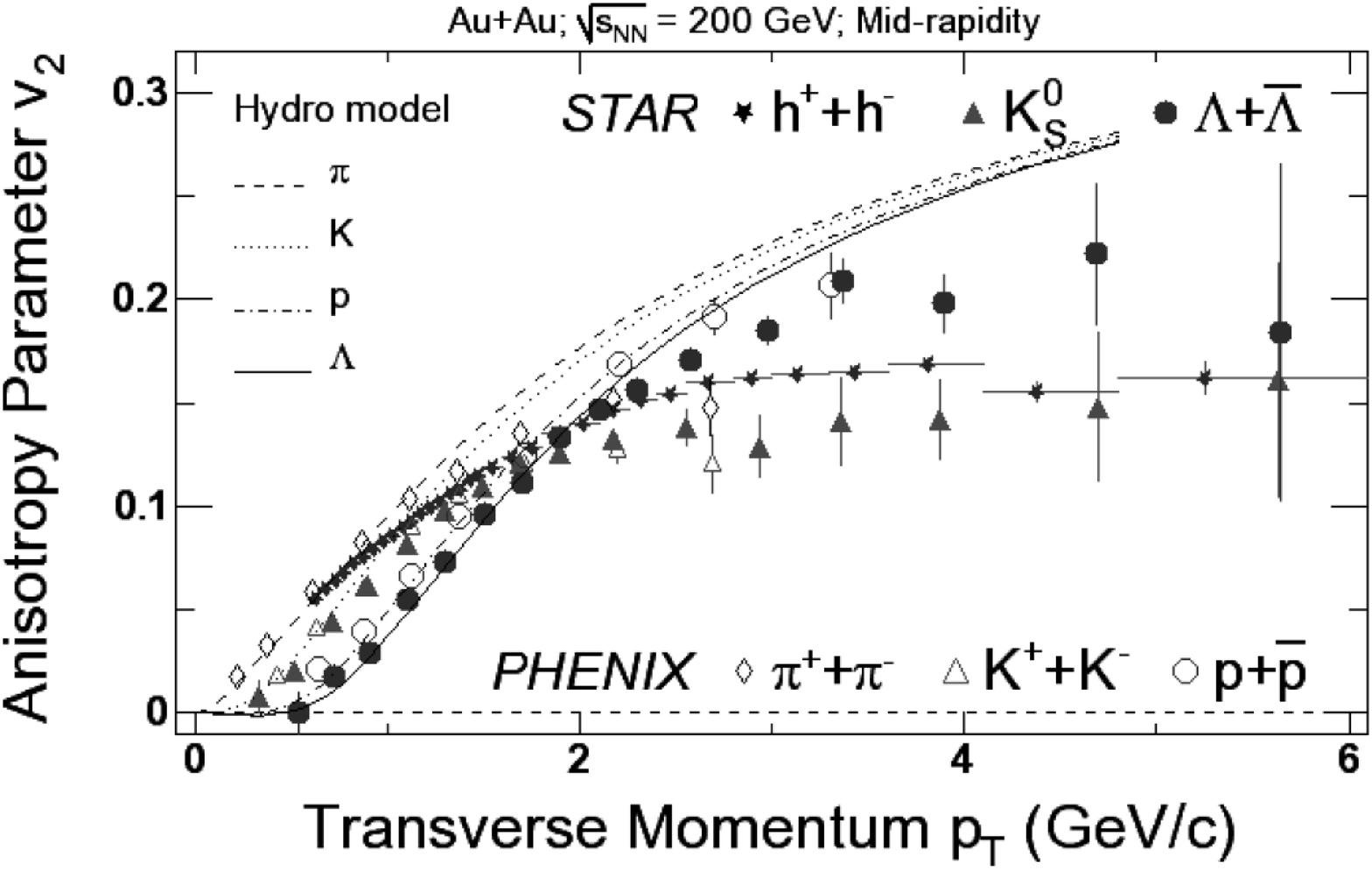}
\end{center}
\caption{Elliptic flow at RHIC energies as function of transverse momentum,
compared to the simple ideal fluid hydrodynamical calculation. Figure
adapted from the presentation of M. Gyulassy, see this volume}
\end{figure}

In addition to the behavior of the elliptic flow phenomena, hydrodynamical
description is found to be very successful in many aspects of SPS and RHIC
data\cite{Review}, thus establishing the general physical scenario of
strongly interacting matter in the process of relativistic heavy ion
collisions. On the other hand, this remarkable success of the hydrodynamics
opens several important questions and understanding of these questions may
lead us to some new physical concepts. In this paper, we discuss some of
these questions.

\section{Relativistic Hydrodynamics}

The hydrodynamical description in nuclear and particle physics is not new%
\cite{Landau}. However, it is also true that to establish a solid
theoretical foundation for the application of hydrodynamical description to
microscopic systems has always been difficult. This is basically because the
starting assumption of the validity of local thermodynamical equilibrium in
hydrodynamics is not at all trivial to be justified for microscopic systems.

For the sake of later discussion, let us here summarize briefly the basic
structure of the relativistic hydrodynamics\cite{Csernai}. Let $T^{\mu \nu }$
be the energy-momentum tensor of the matter. Then the dynamics of the system
should obey the conservation law of this tensor,%
\begin{equation}
\partial _{\mu }T^{\mu \nu }=0.  \label{hydro}
\end{equation}%
For the case of a perfect fluid in local equilibrium, we can write%
\begin{equation}
T^{\mu \nu }=\left( \varepsilon +p\right) u^{\mu }u^{\nu }-pg^{\mu \nu },
\label{Tmunu_ideal}
\end{equation}%
where $\varepsilon ,p$ and $u^{\mu }$ are, respectively, the proper energy
density, pressure and four-velocity of the fluid element. Eq.(\ref{hydro})
should be complemented by the continuity equations for conserved currents
such as the baryon number, 
\begin{equation}
\partial ^{\mu }\left( nu_{\mu }\right) =0,  \label{Baryon}
\end{equation}%
and also the equation of state which describes the thermodynamical
properties of the matter. The equation of state may be specified in the form
of%
\begin{equation}
\varepsilon =\varepsilon \left( n,s\right) ,  \label{e_dens}
\end{equation}%
where $n$ and $s$ are the baryon number and the entropy densities,
respectively. For ideal fluid case, we can also derive the conservation of
entropy, 
\begin{equation}
\partial ^{\mu }\left( su_{\mu }\right) =0.  \label{entropy}
\end{equation}%
from Eqs.(\ref{hydro},\ref{Tmunu_ideal}) and (\ref{Baryon}) using the
Gibbs-Duhem relation,%
\begin{equation}
dp=\mu dn+Tds,  \label{Gibbs}
\end{equation}%
which can be uniquely determined once Eq.(\ref{e_dens}) is specified,
together with the first law of thermodynamics, $d\varepsilon =Tds+\mu dn$
and the Euler relation for extensivity $\varepsilon +p=Ts+\mu n.$

These hydrodynamical equations can also be derived from the action principle%
\cite{variational}. They can be solved numerically to give the space-time
development of all the necessary thermodynamical variables and the fluid
velocity\cite{SPH,Review}.

\section{Some Open Problems of Relativistic Hydrodynamics}

As mentioned in Introduction, the ideal hydrodynamical description for the
dynamics of hot and dense matter observed in RHIC experiments works
amazingly well, especially for the collective flow. The success of the
approach indicates that the very early equilibration of the partonic gas
takes place. The nature of the QGP seems rather a strongly interacting fluid
(sQGP) than a ideal free parton gas\cite{Shuryak-ideal}. On the other hand,
there exist several open problems in the interpretation of data in terms of
the hydro model. These questions require careful examination to extract
quantitative and precise information on the properties of QGP. Furthermore,
one should note that several very different physical scenarios within the
hydrodynamical approach (e.g. the use of continuous emission, sudden
freeze-out and rQMD final state cascade, or some drastically simplified
hydro models such as blast wave solution, etc.) can give rise equally good
results in reproducing the observables with suitable choices of parameters.
In a way, one may say that the hydro signature is \textquotedblleft
robust\textquotedblright , but on the other hand, this could be a synonym of
\textquotedblleft insensitive\textquotedblright . In order to extract more
precise information on the properties of the quark-gluon plasma and the
mechanism of phase transition in the bulk QCD matter, we should clarify
these points and refine the physical parameters used in these approaches.

With respect to this point, one should remind a important point of the
collective flow. When we use the equation of state to describe the pressure
as function of density, the hydrodynamical equations are meaningful only if
the approximation of local thermal equilibrium is valid. On the other hand,
as far as Eq.(\ref{hydro}) is concerned, it is nothing but the local
conservation of energy and momentum density. Therefore, it may be possible
that a system which is completely out of local thermal equilibrium can also
manifest a flow pattern depending on the origin of the force. For example,
as a well-known example, we may recall that the Schr\"{o}dinger equation%
\begin{equation*}
i\hbar \frac{\partial \psi }{\partial t}=-\frac{\hbar ^{2}}{2m}\nabla
^{2}\psi +V\psi 
\end{equation*}%
can also be written just as if the form of hydrodynamics, 
\begin{align*}
\frac{\partial }{\partial t}\vec{v}+\left( \vec{v}\cdot \nabla \right) \vec{v%
}& =-\frac{1}{m}\nabla p_{q}-\frac{1}{m}\nabla V, \\
\frac{\partial n}{\partial t}+\nabla \cdot \left( n\vec{v}\right) & =0,
\end{align*}%
where%
\begin{equation*}
\vec{v}=\frac{1}{m}\nabla \left( \arg \psi \right) ,\ \ \ \ n=\left\vert
\psi \right\vert ^{2}.
\end{equation*}%
Here, 
\begin{equation*}
p_{q}=-\frac{1}{2m}\frac{1}{\left\vert \psi \right\vert }\nabla
^{2}\left\vert \psi \right\vert 
\end{equation*}%
comes from the quantum uncertainty principle and serves as the pressure, but
nothing to do with the thermal equilibrium. In this example, the
\textquotedblleft pressure gradient\textquotedblright comes from the quantum
uncertainty principle. That is, a flow phenomena does not necessarily
indicates the validity of the local thermodynamical equilibrium.

In the following, we list several crucial points of the hydrodynamical
approach to the relativistic heavy ion reactions.

\begin{itemize}
\item {\large Initial conditions}
\end{itemize}

When two nuclei collide, a huge number of partonic degrees of freedom are
excited. At the first instant of the collision, these partonic degrees of
freedom are far from thermal equilibrium. To obtain the initial condition
suitable for the hydrodynamical scenario, we should understand how these
partonic degrees of freedom reach the local equilibrium state, forming the
quark-gluon plasma. It seems that the time scale to achieve the thermal
equilibrium in partonic cascade calculation within reasonable values of
parton cross section is much larger than the time scale required for the
hydrodynamical models. Thus the success of hydrodynamics cast a very
interesting question of how such an early thermalization can be attained in
the initially created QCD partonic excitations. Several new concepts have
been proposed. For example, as the mechanism of quick isotropization from
the initial momentum distribution of partons (which is predominantly
longitudinal), the so-called Weibel instabilities known in plasma physics is
shown to be effective also in the case of QCD\cite{Randrup,PlasmaInst,Arnold}%
.

However, the isotropization itself is a necessary but not a sufficient
condition for the thermal equilibrium. With respect to the mechanism of
thermalization, another interesting notion, called pre-thermalization, is
recently developed. In a scalar field theoretical model, it is shown that
when the system is set to an excited state, the validity of equation of
state (the functional relation between energy density and pressure) is
attained well before the real thermal equilibrium is reached\cite{Pretherm}.
With respect to this point, we have discussed a possible relation between
the mechanism of early thermalization to the non-extended statistical
mechanics\cite{NonExt}. Furthermore, the approach from the color glass
condensate may provide an initial momentum distribution easy to achieve a
quick thermalized state of partons because of saturation mechanism. Several
investigations based on the CGC approach to calculate the initial energy
distribution have been carried out\cite{Hirano}.

\begin{itemize}
\item {\large Event-by-event fluctuations.}
\end{itemize}

Even the mechanism of thermalization is fast enough, the initial condition
attained by high energy nuclear collision event is far from smooth as shown
by the event generators such as NeXUS code\cite{EvenGen}. Furthermore, this
initial configuration fluctuates collision by collision very largely, even
for central collision of heavy nuclei. It has been pointed out that the
effect of even-by-event fluctuations due to the different initial conditions
are crucial for quantitative studies of observables such as elliptic flow%
\cite{Flut}. Since the dynamical evolution of hydrodynamics is extremely
nonlinear, the event average of any hydrodynamical obserbable is quite
different from the value calculated from the averaged smooth initial
configuration, which is the hydrodynamcal calculations commonly used. The
fluctuation of $v_{2}$ coefficient must contain the dynamical information of
the state formed in the early stage of the collision.

\begin{itemize}
\item {\large Finite size effect}
\end{itemize}

Strictly speaking, the hydrodynamics is essentially the zero mean-free path
(or correlation length) approximation for the microscopic degrees of freedom
compared to the system size. For the nuclear collisions, this is not a
trivial approximation. As we know that the nuclear ground state has finite
surface thickness (of the order of 3 fm) but if we try to describe the
density distribution in terms of zero mean free-path hydrodynamics, we would
have zero surface thickness for the ground state mass distribution. The
parameter which characterizes the degree of approximation with respect to
the finite size of the system would be%
\begin{equation*}
x=\frac{\lambda _{micro}}{\lambda _{Hydro}}
\end{equation*}%
where $\lambda _{micro}$ is the typical mean-free path (or the correlation
length) of the constituent particles and $\lambda _{Hydro}$ is the
hydrodynamical inhomogeneity scale, which is typically%
\begin{equation}
\lambda _{Hydro}\simeq \left\vert \frac{1}{\rho }\nabla \rho \right\vert
^{-1}.  \label{L_hydro}
\end{equation}%
Hydrodynamics is valid for $x\ll 1$, so that when the local inhomogeneity
scale in the system is of the order of the mean-free path, the hydrodynamics
should break down.

In order to take into account for the finite $x$ values, as relativistic
extension of the Weizs\"{a}cker term, we may introduce the gradient term in
the action of hydrodynamics as 
\begin{equation*}
I=\int d^{4}x\ \left[ -\varepsilon \left( n,s\right) +\frac{1}{2}\sigma
\left( \nabla ^{\mu }n\right) ^{2}\right] ,
\end{equation*}%
where $\sigma $ is a quantity related to the surface tension in the static
limit and $\nabla ^{\mu }$ is the four-gradient projected to the hyperplane
perpendicular to the velocity,%
\begin{equation*}
\nabla ^{\mu }=\left( g^{\mu \nu }-u^{\mu }u^{\nu }\right) \partial _{\nu
}=P^{\mu \nu }\partial _{\nu }.
\end{equation*}%
where $P^{\mu \nu }=$ $\left( g^{\mu \nu }-u^{\mu }u^{\nu }\right) $ is the
projection operator to the hypersurface perpendicular to the four-velocity $%
u^{\mu }$. The general energy-momentum conservation can be obtained from the
N\"{o}ther theorem and the resulting equation of motion (for constant $%
\sigma $ case) becomes%
\begin{equation*}
\partial _{\mu }\widetilde{T}^{\mu \nu }=-\partial _{\mu }\left( \sigma
(\nabla ^{\mu }n)(\partial ^{\nu }n)-\sigma \dot{n}u^{\mu }\nabla ^{\nu }n-%
\frac{\sigma }{2}g^{\mu \nu }(\nabla ^{\tau }n)(\partial _{\tau }n)\right) 
\end{equation*}%
where 
\begin{equation*}
\widetilde{T}^{\mu \nu }=\varepsilon u^{\mu }u^{\nu }-P^{\mu \nu }\widetilde{%
p}
\end{equation*}%
with 
\begin{equation*}
\widetilde{p}=p+\sigma n\partial _{\nu }(\nabla ^{\nu }n).
\end{equation*}

Unfortunately, the above equation becomes dynamically much more complicated
than the usual hydrodynamics. In any case, the finite size effect on the
local energy density is an important question, in particular with respect to
the properties of phase transition. A further investigation on this
direction is now in progress\cite{FiniteV}.

\begin{itemize}
\item {\large Viscosity effects.}
\end{itemize}

The comparison of flow phenomena to the ideal fluid calculation indicates
that the emergence of new state of matter which flows like an ideal-fluid at
RHIC energies. This idea that the QGP behaves a real ideal fluid raised an
interesting question when the viscosity for the strong coupling limit of 4D
conformal theory obtained from the supersymmetric string theory found to be
a very small\cite{Son-viscosity}. On the other hand, Hirano and Gyulassy
argue that this is due to the entropy density of the QGP which is much
larger than the hadronic phase\cite{Miklos}. To be precise, a quantitative
and consistent analysis of the viscosity within the framework of
relativistic hydrodynamics has not yet been done completely. This is because
the introduction of dissipative phenomena in relativistic hydrodynamics
casts difficult problems, both conceptual and technical. Several works have
been done in this direction\cite{Viscous}. We will discuss later this point
more in detail in the next section.

\begin{itemize}
\item {\large Final hadron spectra from the hydrodynamical model.}
\end{itemize}

To analyze the physical observables in terms of the hydrodynamical scenario,
we have to construct the particle spectra from the hydro solution. As the
hydrodynamical expansion proceeds, the fluid becomes cooled down and
rarefied, thus leading to the \textit{decoupling} of the constituent
particles. At this stage, these particles do not interact any more and
free-stream to the detector. Long-lived resonances and other unstable
particles may decay on the way to the detector after this instant of
decoupling phase. In the standard hydrodynamical models, one introduces the
concept of freeze-out, which assumes that particle emission occurs on a
sharp three-dimensional surface (defined for example by the local
temperature, $T(x,y,z,t)=$ constant). Before crossing it, particles have a
hydrodynamical behavior, and after they free-stream toward the detectors,
keeping memory of the conditions (flow, temperature) of where and when they
crossed the three dimensional surface. The Cooper-Frye formula \cite{co74}
gives the invariant momentum distribution in this case 
\begin{equation}
Ed^{3}N/dp^{3}=\int_{\sigma }d\sigma _{\mu }p^{\mu }f(x,p).  \label{CF}
\end{equation}%
$d\sigma _{\mu }$ is the surface element 4-vector of the freeze out surface $%
\sigma $ and $f$ the thermal distribution function of the type of particles
considered. The space-time dependence comes from those of thermodynamical
parameters, such as temperature and chemical potential. This is the formula
implicitly used in all standard thermal and hydrodynamical model
calculations.

Although simple and elegant, this sudden freeze-out is not only an
idealization but also contains some problems such as conservation of energy
and momentum, negative flux and artificial entropy production\cite{Csernai}.
To remedy this approach several approaches have been proposed such as the
use of URQMD code coupled to the final state of the hydrodynamics\cite%
{br99,Stoecker}. However, usually these calculations are rather complex and
time consuming in practice.

The most important effect which should effect the form of particle spectra
is that not every particles are emitted from the same hypersurface specified
by a unique temperature. The continuous emission approach\cite{gr} still
uses the equilibrium momentum distribution of particles but they can be
emitted continuously during the hydrodynamical evolution of the system. It
also takes into account the absorption effects while the emitted particle
from the inside traverses the surrounding hadronic matter. This accounts for
the finite size of the system in the final particle spectra. In contrast to
the usual sudden freeze-out, it is found that the final hadrons can be
emitted from a broad range of temperatures. It should be emphasized that the
two extremely cases, a sharp temperature surface of sudden freeze-out and
almost flat distribution of temperature of the continuous emission scenario
give equally good description of the observed spectra and flow\cite{Review}.

\section{Viscosity and Causality}

To extract more precise quantitative conclusion on the ideal nature of the
QGP fluid, we should study the effect of dissipative processes on the
collective flow variables. However, a covariant theory of dissipative
phenomena is not trivial at all. We know that the diffusion equation is not
covariant, and the equation is parabolic so that the propagation of signal
has an infinite velocity. Landau introduced the dissipative effects in the
relativistic hydrodynamics, but it is known that the formalism of Landau\cite%
{Landau} of relativistic viscous fluid still leads to the problem of
acausalily. Relativistic covariance is not the sufficient condition for a
consistent relativistic dissipative dynamics. To cure this problem, the
second order thermodynamics was developed by Israel, Stewart and Miller\cite%
{Israel}. However, this theory is too general and complex, containing many
unknown parameters which make difficult the application of the theory to
practical problems. Furthermore, the theory contains the third order time
derivatives, introducing additional difficulties of initial value problem
and numerical procedure. In our opinion, this is not the unique approach to
the consistent relativistic dissipative hydrodynamics. Here, we show that an
alternative theory which satisfies the minimum conditions of covariance and
causal propagation of signal, and in the small viscosity limit, it recovers
the usual ideal hydrodynamics.

To this, we first note that the problem of acausal propagation in usual
diffusion equation,%
\begin{equation*}
\frac{\partial }{\partial t}n=-\zeta \nabla ^{2}n
\end{equation*}%
can be cured by the introduction of the relaxation time as%
\begin{equation}
\tau _{relax}\frac{\partial ^{2}n}{\partial t^{2}}+\frac{\partial n}{%
\partial t}=\zeta \nabla ^{2}n,  \label{tele}
\end{equation}%
converting the parabolic equation to the hyperbolic equation. For a suitable
choice of parameters, $\tau _{relax}$, we can recover the causal propagation
of the diffusion flux\ref{MorseFeshbach}. Physically this can be understood
as following. In general, the diffusion equation is the combination of two
equations,

\begin{itemize}
\item Continuity equation,%
\begin{equation}
\frac{\partial n}{\partial t}+\nabla \cdot \vec{j}=0,  \label{Cont}
\end{equation}

\item Generation of the irreversible current due to the thermodynamical
force,%
\begin{equation}
\vec{j}=-L\nabla \frac{\delta F\left( n\right) }{\delta n}  \label{Current}
\end{equation}%
and within the linear response of the system, we have%
\begin{equation*}
\vec{j}=-\zeta \nabla n
\end{equation*}%
where $F$ is the thermodynamical potential and $\zeta $ is in general a
function of thermodynamical quantities, but here we assume to be constant
for the sake of simplicity.
\end{itemize}

The origin of acausality resides in Eq.(\ref{Current}) than the continuity
equation, (\ref{Cont}), since this means that the action of the force $%
\nabla \left( \delta F/\delta n\right) $ immediately generates the physical
current $\vec{j}.$ We may think of relaxation phenomena for this process.
This can be done phenomenologically by introducing the retardation function 
\begin{equation}
G\left( t,t^{\prime }\right) =\frac{1}{\tau _{relax}}e^{-\left( t-t^{\prime
}\right) /\tau _{relax}}  \label{relax}
\end{equation}%
and rewrite Eq.(\ref{Current}) as%
\begin{equation*}
\vec{j}=-\int_{-\infty }^{t}G\left( t,t^{\prime }\right) \nabla F\left( \vec{%
r},t^{\prime }\right) dt^{\prime }.
\end{equation*}%
In the limit of $\tau _{relax}\rightarrow 0,$ we have $G\left( t,t^{\prime
}\right) \rightarrow \delta \left( t-t^{\prime }\right) $ so that this
recovers the original equation, (\ref{Current}). Now%
\begin{equation*}
\frac{\partial \vec{j}}{\partial t}=-\frac{1}{\tau _{relax}}\nabla F-\frac{1%
}{\tau _{relax}}\vec{j}.
\end{equation*}%
Taking the time derivative of Eq.(\ref{Cont}) and substituting the above
equation,%
\begin{eqnarray*}
\frac{\partial ^{2}n}{\partial t^{2}} &=&\nabla \cdot \left( \frac{1}{\tau
_{relax}}\nabla F+\frac{1}{\tau _{relax}}\vec{j}\right) \\
&=&\frac{1}{\tau _{relax}}\left( -\frac{\partial n}{\partial t}+\nabla
^{2}F\right)
\end{eqnarray*}%
obtaining Eq.(\ref{tele}). Eq.(\ref{tele}) is usually referred to as
telegraphic equation. For more microscopic derivation of telegraphic
equation, see \cite{Jou,Koide1,Koide2}.

We use the above mechanism to derive the relativistic dissipative
hydrodynamics to transform the Landau formulation to satisfy causality. For
this purpose, first let us review briefly the essential part of the Landau
derivation of the relativistic dissipative hydrodynamics. Landau requires
the conservation laws,

\begin{eqnarray}
\partial _{\mu }T^{\mu \nu } &=&0,  \label{TmunuConserve} \\
\partial _{\mu }N^{\mu } &=&0.  \label{NConserve}
\end{eqnarray}%
In the presence of dissipative phenomena, the energy momentum tensor $T^{\mu
\nu }$ and the baryonic current $N^{\mu }$ are not given by Eqs.(\ref%
{Tmunu_ideal}) (\ref{Baryon}) anymore, but instead,%
\begin{equation}
T^{\mu \nu }=\varepsilon u^{\mu }u^{\nu }-P^{\mu \nu }\left( p+\Pi \right)
+\pi ^{\mu \nu },  \label{NeTmunu}
\end{equation}%
\begin{equation}
N^{\mu }=nu^{\mu }+\nu ^{\mu },  \label{NweN}
\end{equation}%
where $\Pi $ and $\pi ^{\mu \nu }$ are the bulk and shear viscous stresses,
respectively and $\nu ^{\mu }$ is the diffusion current of baryon number.
For these, we require the constraints, $u_{\mu }\pi ^{\mu \nu }=0$ and $%
u_{\mu }\nu ^{\mu }=0.$ With these terms, of course the conservation of
entropy Eq.(\ref{entropy}) is not valid anymore, instead we have

\begin{equation}
\partial _{\mu }\left( su^{\mu }-\alpha \nu ^{\mu }\right) =\frac{1}{T}%
\left( -P^{\mu \nu }\Pi +\pi ^{\mu \nu }\right) \partial _{\mu }u_{\nu }-\nu
^{\mu }\partial _{\mu }\alpha ,  \label{s-current}
\end{equation}%
where $\alpha =\mu /T$. Landau identifies the term%
\begin{equation}
\sigma ^{\mu }=su^{\mu }-\alpha \nu ^{\mu }  \label{sigma}
\end{equation}%
as the entropy current and requires the positive definiteness of the
right-hand side of Eq.(\ref{s-current}) as

\begin{equation}
\frac{1}{T}\left( -P^{\mu \nu }\Pi +\pi ^{\mu \nu }\right) \partial _{\mu
}u_{\nu }-\nu ^{\mu }\partial _{\mu }\alpha \geq 0,  \label{Landau}
\end{equation}%
obtaining the following expressions for the viscous terms,%
\begin{equation}
\Pi =\zeta \partial _{\alpha }u^{\alpha },\ \ \ \widetilde{\pi }_{\mu \nu
}=\eta \partial _{\mu }u_{\nu },\ \ \ \widetilde{\nu }_{\mu }=-\kappa
\partial _{\mu }\alpha   \label{viscous}
\end{equation}%
and introduce the projection operator to satisfy the orthogonality condition
for the viscous term to the fluid velocity as 
\begin{equation}
\pi ^{\mu \nu }=P^{\mu \alpha \nu \beta }\widetilde{\pi }_{\alpha \beta },\
\ \nu ^{\mu }=P^{\mu \alpha }\widetilde{\nu }_{\mu },  \label{projection2}
\end{equation}%
where $P^{\mu \alpha \nu \beta }$ is the double symmetric traceless
projection,%
\begin{equation}
P^{\mu \alpha \nu \beta }=\frac{1}{2}\left( P^{\mu \alpha
}P^{\nu \beta }+P^{\nu \alpha }P^{\nu \nu }\right) -\frac{1}{3}P^{\mu \nu }P^{\alpha
\beta }.
\end{equation}

As mentioned, this Landau scheme leads to the acausal propagation of thermal
current. However, at this moment, generalization of these equation in order
to obtain to hyperbolic equations is self-evident. We introduce the
retardation integral in each viscous term before the projection,

\begin{eqnarray}
\Pi  &=&-\int^{\tau }d\tau ^{\prime }G\left( \tau ,\tau ^{\prime }\right)
\zeta \partial _{\alpha }u^{\alpha }\left( \tau ^{\prime }\right) ,
\label{Integral01} \\
\tilde{\pi}^{\mu \nu } &=&\int^{\tau }d\tau ^{\prime }G\left( \tau ,\tau
^{\prime }\right) \partial ^{\mu }u^{\nu }\left( \tau ^{\prime }\right) ,
\label{Integral02} \\
\widetilde{\nu }^{\mu } &=&-\int^{\tau }d\tau ^{\prime }G\left( \tau ,\tau
^{\prime }\right) \kappa \partial ^{\mu }\alpha \left( \tau ^{\prime
}\right) ,  \label{Integral03}
\end{eqnarray}%
where $\tau $ is the local proper time. These integrals are equivalent to
the differential equations%
\begin{eqnarray}
\Pi  &=&-\zeta \partial _{\alpha }u^{\alpha }+\gamma \frac{d\Pi }{d\tau }
\label{derivative01} \\
\tilde{\pi}^{\mu \nu } &=&\eta \partial ^{\mu }u^{\nu }-\gamma \frac{d\tilde{%
\pi}^{\mu \nu }}{d\tau }  \label{derivative02} \\
\widetilde{\nu }^{\mu } &=&-\kappa \partial ^{\mu }\alpha +\gamma \frac{d%
\widetilde{\nu }^{\mu }}{d\tau }  \label{derivative03}
\end{eqnarray}%
with 
\begin{equation*}
\frac{d}{d\tau }=u^{\mu }\partial _{\mu }
\end{equation*}%
is the total derivative with respect to the proper time. These equations,
after the projection (Eq.\ref{projection2}), can be compared to the
Israel-Stewart form,

\begin{eqnarray}
\Pi &=&-\zeta \left( \partial _{\alpha }u^{\alpha }-\beta _{_{0}}\frac{d\Pi 
}{d\tau }-\alpha _{_{0}}\partial _{\alpha }\nu ^{\alpha }\right)
\label{Israel01} \\
\pi ^{\mu \nu } &=&\eta P^{\mu \alpha \nu \beta }\left( \partial _{\alpha
}u_{\beta }-\beta _{2}\frac{d\pi ^{\mu \nu }}{d\tau }-\alpha _{_{1}}\partial
_{\alpha }\nu _{\beta }\right)  \label{Israel02} \\
\nu ^{\mu } &=&-\kappa p^{\mu \nu }\left( \partial _{\nu }\alpha -\beta _{1}%
\frac{d\nu ^{\nu }}{d\tau }+\alpha _{_{0}}\partial _{\nu }\Pi +\alpha
_{_{1}}\partial _{\alpha }\pi _{\nu }^{\alpha }\right) .  \label{Israel03}
\end{eqnarray}%
They are similar, but we can see that the Israel-Stewart form contains the
more general linear combination of the second order variables. However, the
most important and essential difference is that, in our formalism, the
projection operators $P^{\mu \nu }$ enter after the integration of these
equations. All the dissipative terms are expressed explicitly in terms of
independent variables of the usual hydrodynamics. In the Israel-Stewart
form, due to the projection operators, integral form as Eqs.(\ref{Integral01}%
,\ref{Integral02},\ref{Integral03}) can not be obtained explicitly.

Our approach has several advantages to the Israel-Stewart formalism. First
of all, we keep the simple physical structure of Landau formalism, but at
the same time we cured its causality problem by introduction of the
relaxation integral, Eqs.(\ref{Integral01},\ref{Integral02},\ref{Integral03}%
). Use of these integral expressions also eliminates the problem of higher
derivatives in time. We only need the past values of the independent
variables to solve numerically. This also eliminates the problem of extra
initial conditions, too. The integral expressions are easy to be evaluated
when we use the Lagrangian coordinate system such as SPHERIO\cite{SPH,Review}%
. Because of the simple form of viscous terms, incorporation of these
equations to the realistic hydro-code such as SPHERIO is relatively easy. A
work on this line is in progress.

\section{Summary}

The hydrodynamics is found to be a very successful tool for the description
of the relativistic heavy ion collisions. However, from the quantitative
point of view, the present hydrodynamical approach still contains many
uncertainties and also some conceptual problems. In this paper, we call
attention to these questions, and discussed some of problems in detail. In
particular, we propose an alternative theory to the Israel-Stewart second
order thermodynamics, where viscous terms are given by the integral
expressions which take into account of the relaxation time. In this way, the
problem of causality is avoided and at the same time a simple physical
structure of Landau formulation has been kept. Other questions as early
thermalization, finite size effects, fluctuations in initial conditions,
etc. should also be studied more in detail When these questions are
clarified, we will have much more detailed knowledge about the dynamics and
properties of the new states of strongly interacting matter. It is also
quite important to look for signals of genuine hydrodynamics, such as shock
wave propagation\cite{Shock-Horst}.

\bigskip

This work is dedicated to Prof. W. Greiner, one of the pioneers of the
subject, on the occasion of his 70th birthday. The authors express their
thanks to H. M. Kiriyama, F. Grassi, Y. Hama, C.E. Aguiar and E. Fraga for
stimulating discussions and kind help. T.Kodama is also grateful for the
kind hospitality of Prof. H. St\"{o}cker during his stay in Frankfurt. This
work was partially supported by FAPERJ, FAPESP, CNPQ and CAPES.

\end{document}